\documentclass[aps,prl,twocolumn,groupedaddress]{revtex4-1}
\usepackage{color} 
\usepackage{xcolor}
\usepackage{graphicx} 
\usepackage{epsfig}
  \usepackage{kpfonts}

\newcommand{\bee}{\begin{equation}}
\newcommand{\ene}{\end{equation}}
\newcommand{\beea}{\begin{eqnarray}}
\newcommand{\enea}{\end{eqnarray}}

\begin{document}
\title{Ion Vortex Beam}
 \author{Chandrasekhar Shukla }
%
 \email{chandrasekhar.shukla@gmail.com}
 \author{Amita Das}
 \affiliation{Institute for Plasma Research, HBNI, Bhat, Gandhinagar - 382428, India }
 
\date{\today}
\begin{abstract} 
Recent advances in intense short pulse laser interaction with thin foil enable us to accelerates the ions upto relativistic velocity in a controlled way. The accelerated ion beams can be utilized to generate the ion vortex beam in a similar way as it has been done for electron, neutron and photon vortex beam case. We present a theoretical  demonstration of ion vortex beam with orbital angular momentum (OAM) generated from a relativistic laser-matter interaction  rearranged into vortex beam by using phase shifter or computer generated holography. We show that a  Laguerre-Gaussian beam exit for Schr{\"o}dinger, Klein-Gordon and Dirac particles. This study will motivate to  explore the existence of ion vortex beam experimentally.  
\end{abstract}
\pacs{} 
 \maketitle 
{\it Introduction.-}
The vortex beam of charged particles is a recent hot topic of interest from both fundamental as well as application point of view\cite{Birula_2000, Bliokh,Lloyd,Uchida,Verbeeck}.  In particle vortex beam, like an optical vortex\cite{Beth,Allen,Neil}, spiraling wave fronts exit which give the orbital angular momentum around the propagation axis. The electron beam is converted into vortex beam by various technique, for example,  using a helical phase plate\cite{Uchida}, computer generated hologram\cite{Verbeeck} etc. A lot of experimental and theoretical work has been done to produce the electron vortex beam in a controlled way.  In recent, the ion beams up to the order of 100s MeV/nucleon in $\mu$m distance have been successfully achieved by irradiating an intense
 short pulse laser (I$\geq 10^{18}$w/$cm^{2}$, 30fs-10ps)  on thin foil/ultra-thin nanofoil of $\mu$m/nm size \cite{Macchi,Daido, Scullion}. This scheme has significant features comparable to conventional
 ion acceleration scheme (RF scheme) in which ion accelerated upto MeV/nucleon in m
 distance. Laser driven ion acceleration scheme have low emittance($\leq$ 0.004 mm-mrad for the transverse emittance and $\leq$ 10-4 eV-s for the longitudinal emittance)\cite{Schnurer,Cowan}, compact ($\sim$10 $\mu$m) and ultra-short duration($\sim$ps) ion bunch comparable to convention ion acceleration scheme.\\ 
We propose an idea to make proton microscopy based on proton vortex beam. In this scheme the proton beams will be generated by laser radiated thin foil. The beam will travel through magnetic field or negative electric field  so that the electrons that are much lighter (low mass$\sim$1/1836) compare to proton beam deflected and only proton beam pass through this magnetic field or electron will be stop by negative electric field while ion  will pass.  After this stage,  proton beam vortex can be   generated by using helical phase plate or computer generated hologram  similar to electron or neutron\cite{Charles} vortex beam generation technique.The schematic diagram of the proposed idea has been shown in Figure 1. The  proton vortex beam  has  some advantage over electron beam for imaging. Since proton have heavier mass compare to electron, they do not deflect much. Therefore the 3-D image by proton vortex beam is better than electron vortex beam. The same technique can be also used to produce the heavy charge ions vortex beam. Recently, it is observed that the rotating atoms can cure the cancer diseases more efficiently\cite{Victor}. The controlled production of proton and or heavy charge ion vortex beam might also give the better result in cancer therapy compare to normal ion beam therapy\cite{Durante, Bin}.  \\
 We investigate the theoretically the quantum behaviour of ion vortex beam in non-relativistic and relativistic in both limit. we assume that ion is fully striped out of electrons and we do not consider the subatomic level of ion (nucleus).   \\ 
          \begin{figure}
                \includegraphics[width=0.5\textwidth]{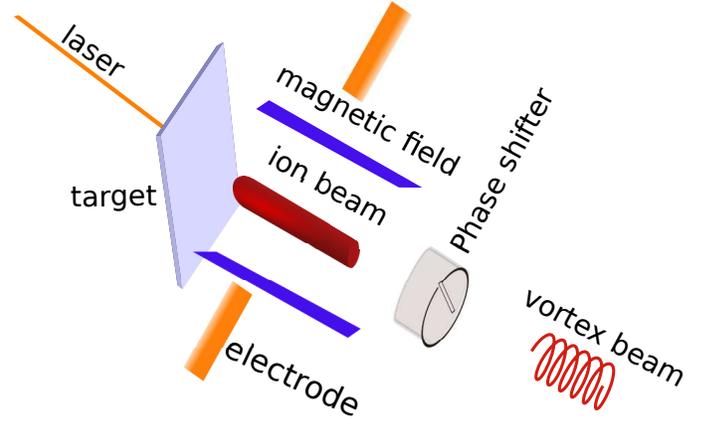} 
                \caption{
                              The schematic diagram of ion vortex beams generation  technique: The laser interacts with thin foil target and generates the ion beams. The ion beam passes through the magnetic field or negative electric field so that the electron which have lighter mass will be deflected or stoped by negative electric field and only ion beams will pass. Then, the ion vortex beam will be generated by using phase shifter or computer generated holograph.}
                \label{fig:ion_vortex1}  
        \end{figure}%

{ \it  Non-relativistic quantum dynamic of monochromatic Laguerre-Gaussian(L-G) beams}- 
 The non-relativistic quantum dynamics of charge particle beam is described by  Schr{\"o}dinger equation
 \begin{equation} 
\left( \frac{\hbar^2}{2m}\bigtriangledown^2+\it i\hbar\frac{\partial}{\partial t}\right)\psi=0
\label{nrel}
\end{equation}\\
We assume the monoenergetic charge particle beam that propagates in z direction ($p_x, p_y \ll p_z$). The Schr{\"o}dinger equation, in cylindrical coordinate system, takes the following form
 \begin{equation} 
 -\frac{\hbar^2}{2m}\left[\frac{1}{\rho}\frac{\partial}{\partial \rho}\left(\rho\frac{\partial}{\partial \rho}\right)+\frac{1}{ \rho^2} \frac{\partial^2}{\partial \phi^2}+\frac{\partial^2}{\partial z^2}\right]\psi=E\psi 
\label{nrel_cly}
\end{equation}\\
The axially symmetric solution of this equation is non-diffracting Bessel beam function\cite{McGloin}, 
\begin{equation}
 \psi^{Bessel}\left(\rho,\phi, z, t \right)=J_{|l|}\left(n\rho\right)exp\left[i\left(l\phi+k_zz-\omega t\right)\right]
\end{equation}
Where $J_{|l|}\left(n\rho\right)$ is first kind Bessel function and  $n$ is transverse wave number. However, integral of probability density which is defined as
\begin{equation}
 P_t=\int P_{|l|}\left(\rho\right)\rho d\rho=\int |\psi|^2 \rho d\rho \propto \int |J_{|l|}\left(n\rho\right)|^2 \rho d\rho
\end{equation}
 divergent in the radial direction in free space and carries non-physical energy (infinite energy)\cite{Bliokh_prx}. Another solution of Schr{\"o}dinger equation in paraxial limit ($\partial^2/\partial^2 z \simeq k^2 + 2ik\partial /\partial z$), which represents the transversely confined vortex beams in free space, are Laguerre-Gaussian (L-G) beams. In this approximation, the L-G
beams have the following form (in natural unit $\hbar=1=c$)\cite{L_Allen},  
\begin{align*}
 \psi^{LG}_{l,n} = \sqrt{\frac{2n!}{\pi(n+|l|)!}}exp\left(-ip^2_zt/2m\right)\frac{(\rho\sqrt{2})^{|l|}}{w^{|l|+1}(z)}exp\left(-\frac{p_z\rho^2}{2(z_R+iz)}\right)\\ \times L^{|l|}_n\left(\frac{2\rho^2}{w^2(z)}\right)e^{il\phi+ikz}exp[-i(2n+|l|+1)tan^{-1}(z/z_R)],
\end{align*}
where $L^{|l|}_n$ are Laguerre polynomials, $n=0,1,2,3,...$ is the radial quantum number, $w(z)=w_0\sqrt{1+z^2/z^2_R}$ is the beam width, and $z_R=p_zw^2(0)/2$ is the Rayleigh range. The average orbital angular momentum of per charge particle is calculated in following way \\
 \begin{equation}
<\hat L_z>=<\psi^{LG}_{l,n}| \hat L_z| \psi^{LG}_{l,n}>= l
\label{OAM}
\end{equation}
it is clear  from eq.(\ref{OAM}) that L-G beams are eigenmodes of oriental angular momentum as well as  eigenmodes of linear momentum at the same time with  eigenvalues  $\hat p \approx  k$.\\
The relativistic quantum dynamics of ion vortex beam whose spin is zero (for example carbon ion $C^{6+}$) is studied by using Klein-Gordon (KG) equation while the relativistic quantum dynamics of proton vortex beam (fermion) whose spin is integral multiple of $ \frac{\hbar}{2}$ is studied by Dirac equation.
\\{ \it Relativistic quantum dynamic of a free boson particle with zero spin}- The Klein-Gordon equation is, 
 \begin{equation}
\left( \frac{1}{c^2}\frac{\partial^2}{\partial t^2}-\bigtriangledown^2+\frac{m^2c^2}{\hbar^2}\right)\psi=0
\label{K-G}
\end{equation}\\
In natural unit $\hbar=1=c$,  it takes the following form, 
\begin{equation} 
\left( \frac{\partial^2}{\partial t^2}-\bigtriangledown^2+m^2\right)\psi=0
\label{K-G}
\end{equation}\\
 The L-G beam is also a solution of KG equation.
 In relativistic regime, the solution of KG equation give  momentum and energy proportional to k and $\omega$ respectively:
\begin{equation}
\nonumber
<\hat p>=<\psi^{LG}_{l,n}| -i\hat \bigtriangledown |\psi^{LG}_{l,n}>= k;
\label{K-G}
\end{equation}
\begin{equation}
<\hat E>=<\psi^{LG}_{l,n}| i\partial/\partial t |\psi^{LG}_{l,n}>= \omega
\label{K-G}
\end{equation}\\
For spin zero particle, the energy-mass-momentum dispersion relation of particle in relativistic regime is 
 \begin{align*} 
E^2=(m^2+p_z^2)\\
E=\pm(m^2+p_z^2)^{1/2}
\label{K-G}
\end{align*}\\
Considering only positive energy solution, the L-G beam takes the following form, 
\begin{align*}
 \psi^{LG}_{l,n} = \sqrt{\frac{2n!}{\pi(n+|l|)!}}exp\left(-iEt\right)\frac{(\rho\sqrt{2})^{|l|}}{w^{|l|+1}(z)}exp\left(-\frac{p_z\rho^2}{2(z_R+iz)}\right)\\ \times L^{|l|}_n\left(\frac{2\rho^2}{w^2(z)}\right)e^{il\phi+ikz}exp[-i(2n+|l|+1)tan^{-1}(z/z_R)],
\end{align*}
The orbital angular momentum of a spin zero particle is conserved i. e.
 \begin{align*} 
[H_0,\hat L^2]=0;
\end{align*}
 where $H_0=-\bigtriangledown^2+m^2$. Therefore, the orbital angular momentum of spin zero particle can be measured separately and the average orbital angular momentum per particle is 
\begin{equation}
<\hat L_z>=<\psi^{LG}_{l,n}| \hat L_z |\psi^{LG}_{l,n}>= l
\end{equation}
{ \it Relativistic quantum dynamic of a free fermion}- The relativistic quantum dynamics of fermion beam is correctly represented by Dirac equation. The Dirac equation has the following form, 
\begin{equation} 
\it{i\hbar{\partial_t}\psi}=\left({\bm{\hat \alpha}\cdot \bf\hat p}+\beta m \right)\it \psi
\label{rel}
\end{equation}\\
where $\psi$ is four-component spinor and  $\bm{\hat\alpha}$ and $\beta$ arethe  basis matrices, defined as,
\begin{equation}
 \bm{\alpha}= 
\left(\begin{array}{cc}
 \sigma  & 0 \\ 0 & \sigma \end{array}\right), \hspace{1cm}
 \beta=
  \left(\begin{array}{cc}
  \bf I  & 0 \\ 0 & \bf -I \end{array}\right),
\end{equation}
$\sigma$ is Pauli matrices and $\bf I$ is the unit matrix.\\
In relativistic quantum dynamics the orbital angular momentum $\hat L=\hat r\times\hat p$ and spin angular momentum $\hat S=\frac{1}{2}\bm{\hat\alpha}$\cite{pauli1,pauli2,pauli3} do not conserve separately i. e. 
\begin{align*} 
[H_0,\hat L]=[{\bm{ \hat\alpha}\cdot {\bf\hat p}+\beta m},\hat r\times\hat p]=-i \bm{\hat\alpha}\times \bf\hat p \neq0;
\end{align*}
\begin{equation} 
[H_0,\hat S]=[{\bm{ \hat\alpha}\cdot {\bf\hat p}+\beta m},\hat S]=i \bm{\hat\alpha}\times \bf\hat p \neq0;
\end{equation}
 However, the total angular momentum($\hat J= \hat L+\hat S$) do conserve i.e.
 \begin{equation} 
[H_0,\hat J]=0;
\end{equation}
  The L-G beam is also a solution of Dirac equation. Since the orbital angular momentum and spin are coupled for Dirac particle, it is not possible to measure the separate current (spin part and orbital part) in vortex beam. Therefore, the existence of vortex is doubted  at fundamental level \cite{Iwo_Birula}. However, The spin operator which has been defined by Pauli is questionable in relativistic regime. There are also others relativistic spin operators suggested which work in specific condition\cite{spin1,spin2,spin3,Foldy, Sud, Pryce}. Only the Foldy–Wouthuysen operator\cite{Foldy, Sud} which is defined as
\begin{equation}
\hat S_{FW}=\frac{1}{2}\hat{\bm \Sigma}+\frac{i\beta}{2\hat p_0}{\bf \hat p\times \bm \alpha}
-\frac{{\bf \hat p\times(\hat{\bm \Sigma}\times \bf \hat p)}}{2\hat p_0(\hat p_0+m_0)}
\end{equation}  
   and the Pryce operator \cite{Pryce} defined as 
   \begin{equation}
\hat S_{Pr}=\frac{1}{2}\hat{\bm \Sigma}+\frac{1}{2}\hat{\bm \Sigma}\cdot\bf\hat p(1-\beta)\frac{\bf \hat p}{\bf {\hat p}^2}
\end{equation}
satisfy the proper relativistic spin operator properties\cite{relativistic_operator}.
The spin operator, orbital angular momentum operator and total angular momentum are conserved separately. Therefore the spin current and orbital angular current can be measured separately. The Foldy-Wouthuysen operator satisfy the following relations,
\begin{equation}
[H_0,\hat S_{FW}]=0;[S_{FW,i}, \hat S_{FW,j}]=i\epsilon_{i,j,k}\hat S_{FW,k},
\end{equation}
The Foldy-Wouthuysen spin operator fails in satisfy proper spin operator in case of electron spin in hydrogen like ion but Pryce operator satisfy the properties of proper spin operator. 
The proton like ion has integral multiple of $ 1/2$ and it has been shown that by using  Foldy-Wouthuysen operator, L-G beam shows the  existence of  relativistic electron (fermion) vortex at fundamental level\cite{Barnett}. Therefore, L-G beam is also a solution of Dirac equation for proton like ion vortex beam.\\
  In conclusion, we demonstrated theoretically that ion beams generated in high intensity laser irradiated ultra thin foil can be used to construct the ion vortex beam with particular angular momentum. This study will motivate 
  to generate ion vortex beam experimentally which can spark the research of ion based microscopy and  fundamental nature of quantum particles.
%

 \bibliographystyle{unsrt}

 \clearpage
\end{document}